\documentclass[12pt]{iopart}
\usepackage{graphics, graphicx, SIunits}

\usepackage{colortbl, array, booktabs, textcomp, multirow, bigdelim, subfigure}
\usepackage[table]{xcolor}

\usepackage{cancel}
\usepackage{geometry}
\definecolor{Gray}{gray}{0.9}
\newcolumntype{M}[1]{>{\centering\arraybackslash}m{#1}}

\begin{document}
\title[Towards the LISA Backlink]{Towards the LISA Backlink: Experiment design for comparing optical phase reference distribution systems}

\author{Katharina-Sophie Isleif$^{1}$, Lea Bischof$^{1}$, Stefan Ast$^{1}$ \footnote{Present address:
Laboratoire ARTEMIS - Observatoire de la C\^ote d'Azur, Boulevard de l'Observatoire, F 06304 NICE Cedex 4, France.}, Daniel Penkert$^{1}$, Thomas S Schwarze$^{1}$, Germ\'an Fern\'andez Barranco$^{1}$, Max Zwetz$^{1}$, Sonja Veith$^{1}$ \footnote{Present address: Leibniz Universit\"at Hannover, Institute for Special Education, Schlo\ss wender Stra\ss e 1, 30159 Hannover, Germany.}, Jan-Simon Hennig$^{1}$ \footnote{Present address: University of Glasgow, Institute for Gravitational Research, University Avenue, Glasgow G12 8QQ, Scotland, UK.}, Michael Tr\"obs$^{1}$, Jens Reiche$^{1}$, Oliver Gerberding$^{1}$, Karsten Danzmann$^{1}$ and Gerhard Heinzel$^{1}$
}
\address{$^1$Leibniz Universit\"at Hannover, Institute for Gravitational Physics and Max Planck Institute for Gravitational Physics (Albert Einstein Institute), Callinstr. 38, 30167 Hannover, Germany \\
 }
\ead{katharina-sophie.isleif@aei.mpg.de}
\vspace{10pt}
\begin{indented}
\item[]\today
\end{indented}

\begin{abstract}
LISA is a proposed space-based laser interferometer detecting gravitational waves by measuring distances between free-floating test masses housed in three satellites in a triangular constellation with laser links in-between. Each satellite contains two optical benches that are articulated by moving optical subassemblies for compensating the breathing angle in the constellation. The phase reference distribution system, also known as backlink, forms an optical bi-directional path between the intra-satellite benches.

In this work we discuss phase reference implementations with a target non-reciprocity of at most $2\pi\,\mathrm{\micro \radian/\sqrt{Hz}}$, equivalent to $1\,\mathrm{pm/\sqrt{Hz}}$ for a wavelength of 1064\,nm in the frequency band from 0.1\,mHz to 1\,Hz. One phase reference uses a steered free beam connection, the other one a fiber together with additional laser frequencies. The noise characteristics of these implementations will be compared in a single interferometric set-up with a previously successfully tested direct fiber connection. We show the design of this interferometer created by optical simulations including ghost beam analysis, component alignment and noise estimation. First experimental results of a free beam laser link between two optical set-ups that are co-rotating by $\pm 1\degree$ are presented. This experiment demonstrates sufficient thermal stability during rotation of less than $10^{-4}\,\mathrm{K/\sqrt{Hz}}$ at 1\,mHz and operation of the free beam steering mirror control over more than 1 week. 
\end{abstract}


Submitted to \CQG

\maketitle
\section{Introduction}
\label{intro}
 The Laser Interferometer Space Antenna (LISA) has recently been chosen by the European Space Agency (ESA) as L3 mission and will consist of three spacecraft (S/C) forming a cartwheel formation following the Earth around the sun \cite{LISAmissionL3,2013DanzmannWhitepaper,Danzmann2003}. LISA will allow us to directly detect gravitational waves of astrophysical sources radiating in the mHz frequency band such as extreme mass ratio inspirals, supermassive black holes and binary neutron stars and to predict potential detections by ground-based gravitational wave detectors, like aLIGO \cite{abbott2016observation} and VIRGO \cite{VIRGOdetect}, many years in advance \cite{Sesana2016}. 

\begin{figure*}
\centering
\subfigure[Sketch of the planned space-based gravitational wave detector LISA. The current design envisages two moving optical benches per spacecraft (S/C) which are, in this case, optically connected by a fiber backlink. \label{fig:LISABL}
]{\includegraphics[width=0.45\textwidth]{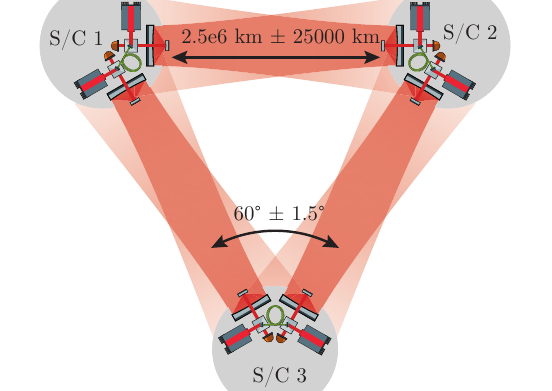}} \qquad
\subfigure[The direct fiber backlink has already been investigated \cite{Fleddermann2017}. Backlink ghost beams are not eliminated, which is why balanced detection and/or a fiber length stabilization, with a high voltage amplifier to drive a ring piezo (PZT) around which the fiber is coiled, is required.]{\includegraphics[width=0.45\textwidth]{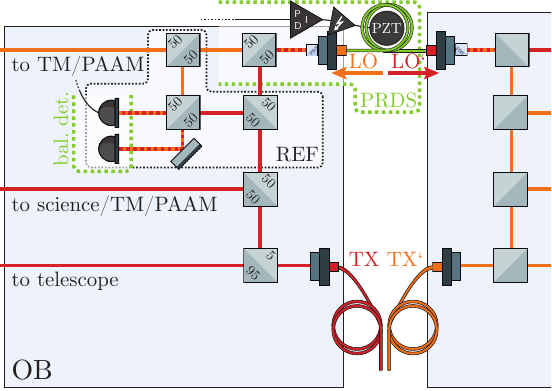}}
\subfigure[A frequency separated fiber link uses an additional local oscillator (ALO) for the intra-satellite bench light exchange and shifts parasitic beats to other heterodyne frequencies.]{\includegraphics[width=0.45\textwidth]{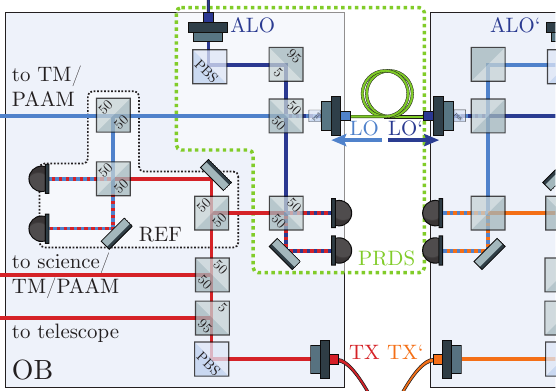}} \qquad
\subfigure[A free beam link completely eliminates the backlink fiber and reduces ghost beams via polarization encoding. Two steering mirrors are used in a control loop to compensate the bench rotation.]{\includegraphics[width=0.45\textwidth]{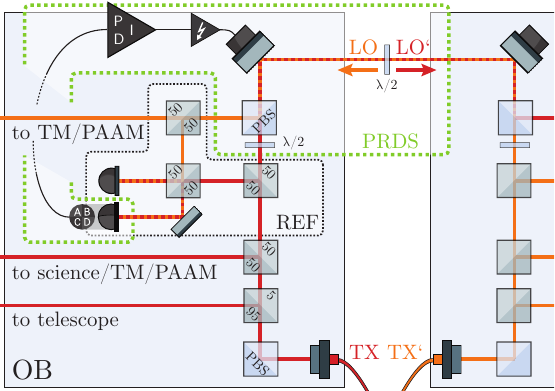}}
\caption{The Laser interferometer space antenna (LISA) (a) requires an optical phase distribution system (PRDS) between two optical benches within one spacecraft, the so-called backlink. Shown here are a direct fiber link in (b), a fiber backlink with frequency separation in (c) and a free beam connection in (d), each scheme integrated on the LISA optical bench. The transmitted (TX) beam of each bench is used as local oscillator (LO) on its counterpart, respectively, enabling heterodyne interferometry in the science, test mass (TM), and point-ahead angle mechanism (PAAM) interferometers as well as in the PRDS.}
\label{fig:LISAand3BL}
\end{figure*}

Figure \ref{fig:LISABL} shows the current inter-satellite interferometry for LISA having an arm-length of $2.5\cdot 10^6\,\mathrm{km}$ with $60\degree$ angles between them. Due to solar system celestial mechanics, the equilateral satellite formation is disturbed. This results in arm breathing, i.e. the arm-length varies by about $\pm 1\%$ and the angles by about $\pm 1.5\degree$ with an annual period. To compensate for the breathing angles the current baseline includes two moving optical subassemblies per S/C, articulating the payload and retaining the interferometric contrast \cite{schutze2014laser}. An alternative approach using a single optical bench and an actuation of the tertiary telescope mirror is also under investigation but not discussed here \cite{LivasInField2017,Brugger2016}. The readout of free-floating test masses is performed relative to the local optical benches similar to LISA Pathfinder interferometry \cite{armano2016sub,heinzel2004ltp}. 

 Laser frequency noise limits the readout of the S/C separations in LISA because it enters the phase measurement due to the arm length mismatches of up to 25000\,km.  This effect is suppressed in data post-processing by time-delay interferometry (TDI) \cite{Tinto05LRR,Otto12CQG}, which forms linear combinations of the local and the inter-S/C phase measurements from different arms with time-delayed versions of themselves. For this purpose a precise phase comparison between the two local lasers on each optical bench within one S/C is required and realized by a phase reference distribution system (PRDS) or backlink.

To be compatible with TDI, the over all phase noise in the backlink is uncritical, but differential (non-reciprocal) phase fluctuations directly enter the measurement and must thus be guaranteed to be small enough to achieve an overall single arm sensitivity of about $10\,\textrm{pm}/\sqrt{\textrm{Hz}}$. Establishing such a bi-directional link requires either a flexible optical connection via fiber or a type of steered free beam set-up that is read out via reference interferometers on each optical bench, achieving a non-reciprocity of $1\,\mathrm{pm/\sqrt{Hz}}$ at frequencies between 0.1\,mHz and 1\,Hz \cite{Steier2009BacksideFiber,Isleif2016Proceed,Fleddermann2017}. Typically, this optical connection is also used to transport laser light from each optical bench to the other to be used as local oscillators for heterodyne interferometry with a few mW of laser power and 5-25\,MHz frequency difference.

The non-reciprocity of a straightforward, direct fiber backlink has already been demonstrated in a previous experiment that compared a fiber connection to a stable reference path \cite{Fleddermann2017}. In addition to post-corrections of angular jitter and subtraction of some unexplained temperature coupling, this experiment required a balanced detection scheme to reduce the phase noise caused by parasitic ghost beams generated by Rayleigh scattering inside the fibers. While an active length stabilization of the backlink fiber and attenuation stages behind the fiber collimators delivering the light onto the bench do reduce the ghost beam coupling, these measures are not sufficient to achieve the required non-reciprocity \cite{Fleddermann2017,Fleddermann2012,Hennig2013}. 

The implementation of this direct fiber-based backlink in LISA is sketched in figure \ref{fig:LISAand3BL}(b). The direct optical connection between the fiber injector optical subassembly (FIOS \cite{killow2016optical}), delivering the transmitted (TX) beam and the local oscillator (LO) beam, leads to back scatter in both fibers and this, in turn, to collinear ghost beams interfering at the critical heterodyne frequencies. These parasitic beats couple into all local interferometers, requiring balanced detection to reduce more than two orders of magnitude of phase noise in each one. 

\section{Comparing ghost beam optimized backlinks}
A direct fiber backlink in combination with balanced detection is able to meet the LISA specifications in a breadboard experiment. However, it is unclear if the in-flight performance can achieve the redundancy implications and the noise characteristics in comparison to alternative schemes.
The noise suppression level of balanced detection required in LISA depends directly on the ghost beam powers which are driven by the amount of fiber back scatter. So far, these back scatter levels have been measured to be on the order of 4\,ppm per meter \cite{Fleddermann2012}, but this is likely to increase in flight due to radiation damage in the fiber core to yet unknown values.
Suppressing more than two orders of magnitude with balanced detection has not yet been demonstrated, and even the use of normalized balanced detection, as shown by Fleddermann et al. \cite{Fleddermann2017}, might be limited by other sources of low-frequency amplitude drifts, leading to additional constrains on the stability, or readout, of the signal amplitudes. 
We intend to experimentally study phase references with less ghost beam coupling in a side-by-side comparison to a direct fiber backlink to better understand their respective limitations and to support the trade-off studies for the LISA optical bench design. Additionally, this experiment is to be conducted in a more representative environment than the previous one by using two optical set-ups rotating against each other. 

The first alternative scheme we will investigate is a fiber-based implementation that uses an additional local oscillator (ALO) frequency to shift the parasitic beat notes generated by back scatter in the backlink fiber to uncritical heterodyne frequencies. Here, no light is sent back into the TX FIOS, such that, in theory, no direct fiber back scatter should spoil the measurements. An additional interferometer provides the phase reference to the ALO which is brought onto the bench by a dedicated FIOS as shown in figure \ref{fig:LISAand3BL}(c). The only known limitation unique to this approach is a double reflection of ghost beams between the ALO and the backlink fiber, which can be suppressed by additional power attenuation. 

The second scheme included here is a free beam connection with active steering control to maintain interferometric contrast and to reduce tilt-to-length coupling. A coupling of the local oscillator light into the TX FIOS is further suppressed by orthogonal polarization encoding of the two counter-propagating free beams. The control of the steering mirrors requires precise tilt sensing, for the purpose of which the reference interferometer is equipped with quadrant photodiodes (QPDs), as shown in figure \ref{fig:LISAand3BL}(d), and read out by means of differential wavefront sensing (DWS) \cite{DWS}. This approach can still be limited by back scatter from the TX fibers, depending on the extinction ratio of the polarization encoding and excess tilt-to-length noise.

\begin{figure*}
\centering
\resizebox{\textwidth}{!}{%
  \includegraphics{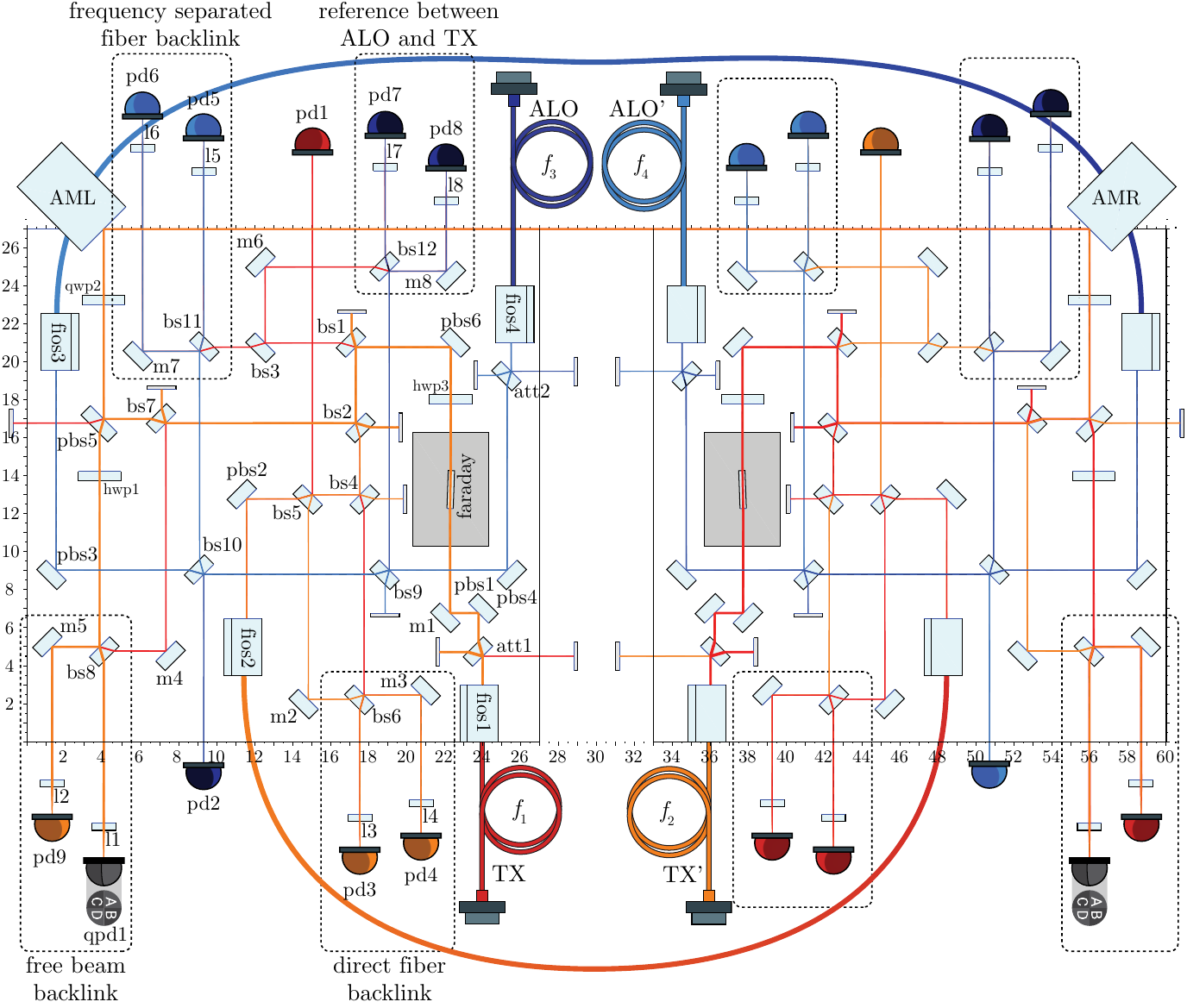}
  }
\caption{Current design of the Three-Backlink Interferometer (TBI). The components and beams on the two optical benches are placed and propagated using an optical simulation tool, the C++ software library IfoCAD. In total, the TBI houses eight monolithic FIOSs ($2 \times$ fios1-fios4), 16 deflection mirrors ($2 \times$ m1-m8), 24 beam splitters ($2 \times$ bs1-bs12) with a splitting ratio of 50\%, four attenuation beam splitters with a reflectivity of 95\% ($2 \times$ att1, att2), two actuator mirrors (AML and AMR), 12 polarizing beam splitters ($2 \times$ pbs1-pbs6 ), 16 lenses ($2 \times$ l1-l8), five half-wave plates ($2 \times$ hwp1,hwp2, $1 \times$ hwp3) (potentially two quarter-wave plates ($2 \times$ qwp1) instead of hwp3), two Faraday rotators, 74 beam dumps, 18 photodiodes ($2 \times$ pd1-pd9) and two quadrant photodiodes ($2 \times$ qpd1). }
\label{fig:3BL}
\end{figure*}

The experiment we designed to compare these schemes with a direct fiber link and to probe their limitations is shown in figure \ref{fig:3BL} and referred to here as Three-Backlink Interferometer (TBI). While arm length mismatches in the interferometers do cause laser frequency noise to leak into their individual measurements, this effect cancels out within each backlink's inter-bench phase sum due to the mirror symmetry of the two optical benches \cite{Fleddermann2017}. Moreover, the differences between the three individual backlinks' phase sums contain only the associated non-reciprocities, providing insight into their unique limitations.
The TBI was planned and optimized with the C++ software library IfoCAD \cite{Wanner12OC,kochkina2013simulating} to be implemented in a quasi-monolithic fashion on ultra-low expansion glass ceramic baseplates with a coefficient of thermal expansion of $1\cdot 10^{-8}\,/\mathrm{K}$. 

The three different backlink schemes are implemented in the TBI using a common TX beam simplifying the comparison. Without additional means the large back scatter from the direct fiber link dominates the noise in all backlinks. To avoid this we integrated not only an attenuator into the common TX beam path after the TX FIOS but also a Faraday isolator. This suppresses the light propagating to the TX FIOS and, hence, the associated ghost beams.
The magnetic field generated by the Faraday is likely exceeding the strict requirements of the gravitational reference sensor, i.e. the test mass, making this implementation unique to the TBI and probably unsuitable for the final LISA optical bench design \cite{DIAZAGUILO201353}.

To optimize the ghost beam suppression in the TBI we used IfoCAD to trace all beams originating from FIOSs or generated at primary as well as secondary surfaces of all optical components and computed which of those reach one of the 20 photodiodes. For each diode we then computed the respective power levels, frequencies and heterodyne efficiencies for the relevant beam combinations. Ignoring beats at uncritical heterodyne frequencies we were able to identify the limiting ghost beam sources \cite{Isleif2016Proceed}. With strategically placed beam dumps in place we find that only ghost beams from fiber back scatter, traveling collinearly to the nominal beam with the same mode, will be limiting. Taking into account a suppression factor of two orders of magnitude for the polarization components which are not included in the IfoCAD simulation, we find that only the interferometer with the direct fiber link scheme will be limited by ghost beams. 

The fundamental noise sources in the TBI, i.e. optical shot noise, electronic noise, and relative intensity noise, depend on the laser power impinging on the photodiodes \cite{trobs2012testing}. Optimizing the power splitting ratios with IfoCAD and choosing an input power of 320\,mW and 160\,mW for the TX and ALO, respectively, we are able to reduce this to negligible levels. The highest resulting path length noise is below $0.018\,\mathrm{pm/\sqrt{Hz}}$, two orders of magnitude below the requirement of $1\,\mathrm{pm/\sqrt{Hz}}$, for an assumed wavelength of $1064\,\mathrm{nm}$ and a photodiode responsitivity of about $0.7\,\mathrm{A/W}$. To monitor, and potentially control, low frequency power fluctuations of each laser the TBI contains dedicated laser power pick-off photodiodes.

The simulation of baseplate rotations around each center reveals a path-length change of $17.02\,\mathrm{mm}$ for a rotation of $\pm 1.5^\circ$. The differential wavefront sensing (DWS) signals calculated by IfoCAD indicate that a decoupling of the remote and far steering mirror tilt-sensing, which is crucial for establishing closed-loop control, can be realized by simply placing focusing lenses in front of the quadrant photo diodes. Simulating a static misalignment of one of the steering mirrors of $10\,\textrm{\textmu rad}$ we find a worst case coupling from steering mirror angle into the interferometer path length on the far bench of $20\,\textrm{\textmu m}/\textrm{rad}$, resulting in a beam tilt noise requirement of $50\,\mathrm{nrad/\sqrt{Hz}}$ to achieve a displacement noise of $1\,\mathrm{pm/\sqrt{Hz}}$. Assuming a DWS coupling coefficient of 4000\,rad/rad this results in a DWS noise requirement of $0.2\,\mathrm{mrad/\sqrt{Hz}}$. Post-corrections can also be applied in case of insufficient closed-loop suppression of the steering mirror control if the DWS sensing noise is below the requirement.

\section{Experimental infrastructure}
\begin{figure*}
\resizebox{\textwidth}{!}{%
  \includegraphics{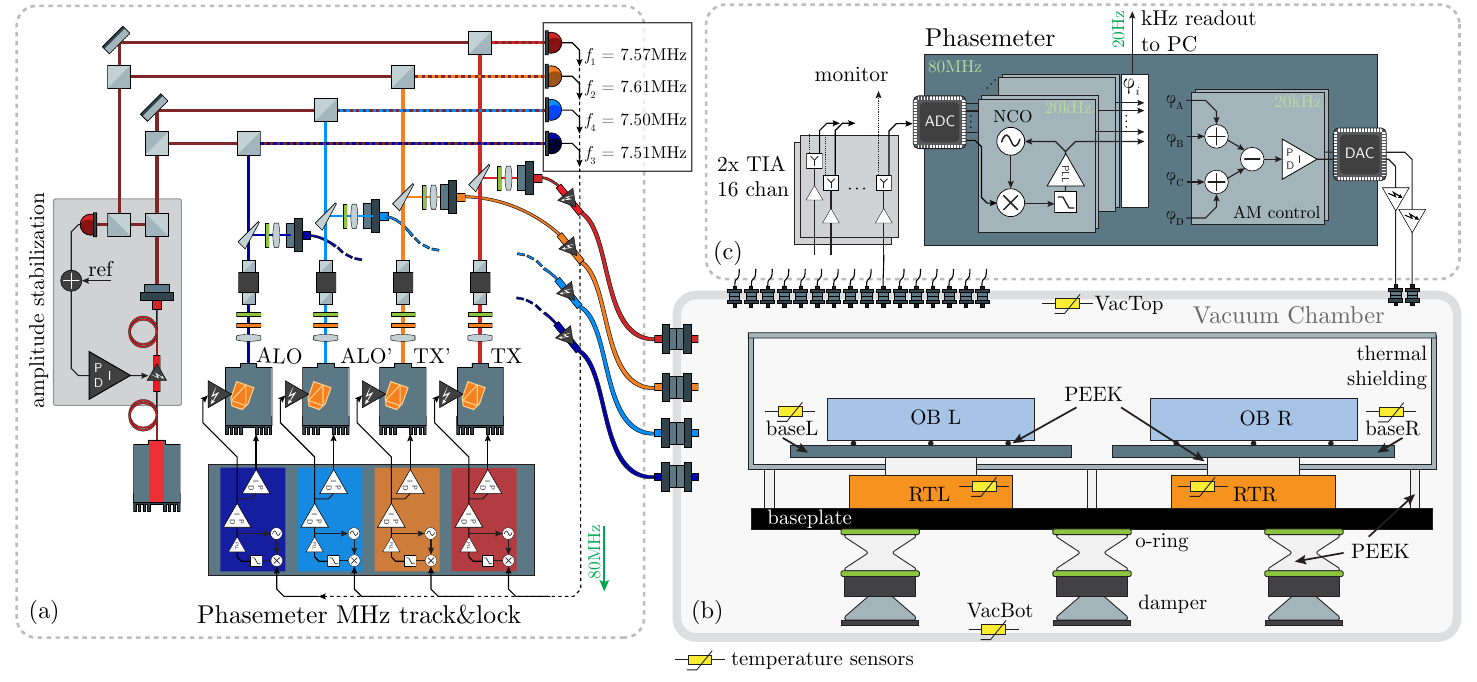}
  }
\caption{The experimental infrastructure depicted in three parts. (a) Laser preparation and offset phase locks. (b) Vacuum chamber, rotation tables and main interferometer. (c) Phase readout and back-end electronics. \\RTL: rotary tables on the left, RTR: rotary table on the right, OB L: optical bench on the left, OB R: optical on the right, TIA: transimpedance amplifier, AM: actuator mirror.}
\label{fig:PreStab}
\end{figure*}

To probe the noise influences in the TBI down to the LISA requirements an experimental infrastructure is required that provides all laser beams, including desired stabilizations, a representative environment, and the phase readout and control electronics. 
We have implemented these parts of the experiment as sketched in figure \ref{fig:PreStab} and in the following we will describe some relevant details.


We supply four single sideband lasers beams to the TBI with constant frequency offsets between them and power levels similar to the values given in section 2. Assuming a maximal arm-length difference between each interferometer on both optical benches of less than 1\,cm the common frequency stability of the lasers should be better than 30\,kHz/$\mathrm{\sqrt{Hz}}$ between 1\,mHz and 1\,Hz. We establish this by offset-phase locking four non-planar ring oscillator (NPRO) lasers, with a nominal wavelength of 1064.5\,nm, to an iodine reference laser with an offset frequency of about 7.5\,MHz. The iodine stabilization provides frequency noise below 100\,Hz/$\mathrm{\sqrt{Hz}}$ at the desired frequencies. The exact offset for each laser is slightly shifted, such that their heterodyne beat notes are at 9.76\,kHz, 39.06\,kHz, 58.59\,kHz, 68.36\,kHz, 97.66\,kHz and 107.42\,kHz. 
 These frequencies are chosen such that the phasemeters, sampling at 80\,MHz, can optimally resolve them and that there are no coinciding harmonics below fourth order. The phase locking is realized digitally with an algorithm implemented in a field-programmable gate array (FPGA) \cite{McNamara2005,diekmann2009analog,gerberding2015readout}. Actuation signals are converted into analog voltages modulating the laser crystals via piezo-electric transducers for fast frequency changes and the crystal temperature for slow changes. The control loops achieve a unity gain frequency of 17\,kHz and a phase margin of 30\degree and they have already been operated continuously over several months. 
 
99\% of the light from each NPRO is fed to the vacuum chamber by 12\,m polarization maintaining fiber feedthroughs. To maintain constant power levels in the TBI we also established analog amplitude stabilizations actuating on fiber-based amplitude modulators, achieving unity gain frequencies of more than 2\,kHz.

The environment to test the TBI has to provide vacuum, a thermal stability of better than $10^{-4}\,\mathrm{K/\sqrt{Hz}}$ at 1\,mHz and two co-rotating baseplates enabling a rotation of about $\pm 1.5\degree$ each. To achieve this we implemented two rotary tables, each one carrying one interferometer, on a common baseplate positioned inside a vacuum chamber. Around the interferometers, excluding the rotary tables, we placed a passive thermal shield covered by multilayer insulation foil. The thermal shield and the common baseplate are thermally isolated from each other as well as from the vacuum chamber by adapters made of polyether ether ketone (PEEK). To reduce an inverted pendulum motion of the rotary table payload we also included vibration damping feet and vibration isolation of the vacuum pumps. To monitor temperature fluctuations we placed thermal sensors throughout the set-up. To reduce the thermal load within the vacuum chamber we minimized the amount of active electronics. Prominently, we use heterodyne frequencies that are low enough to separate photodiodes from trans-impedance amplifiers. 

The phase of the photodiode currents has to be extracted with a precision of better than $2\pi\,\mathrm{\micro \radian/\sqrt{Hz}}$. The data has to be decimated and sent to a PC while some signals have to be used in a realtime implementation of the steering mirror control loops, generating actuation voltages for all degrees of freedom. We use an FPGA-based tracking phasemeter  \cite{gerberding2015readout} with a sampling rate of 80\,MHz to deduce the phase on each photodiode segment after the currents have been converted into a voltage and digitized. The channels containing the free beam DWS information are combined and fed into four controllers, one for each axis of each steering mirror. The resulting actuation signals are converted back into voltages and operate the steering mirror angles via high voltage amplifiers and piezo-electric transducers.

\section{Free beam experiment}

\begin{figure*}
\centering
\subfigure[2D sketch of the free beam link including beam rays simulated with IfoCAD.]{\includegraphics[width=0.75\textwidth]{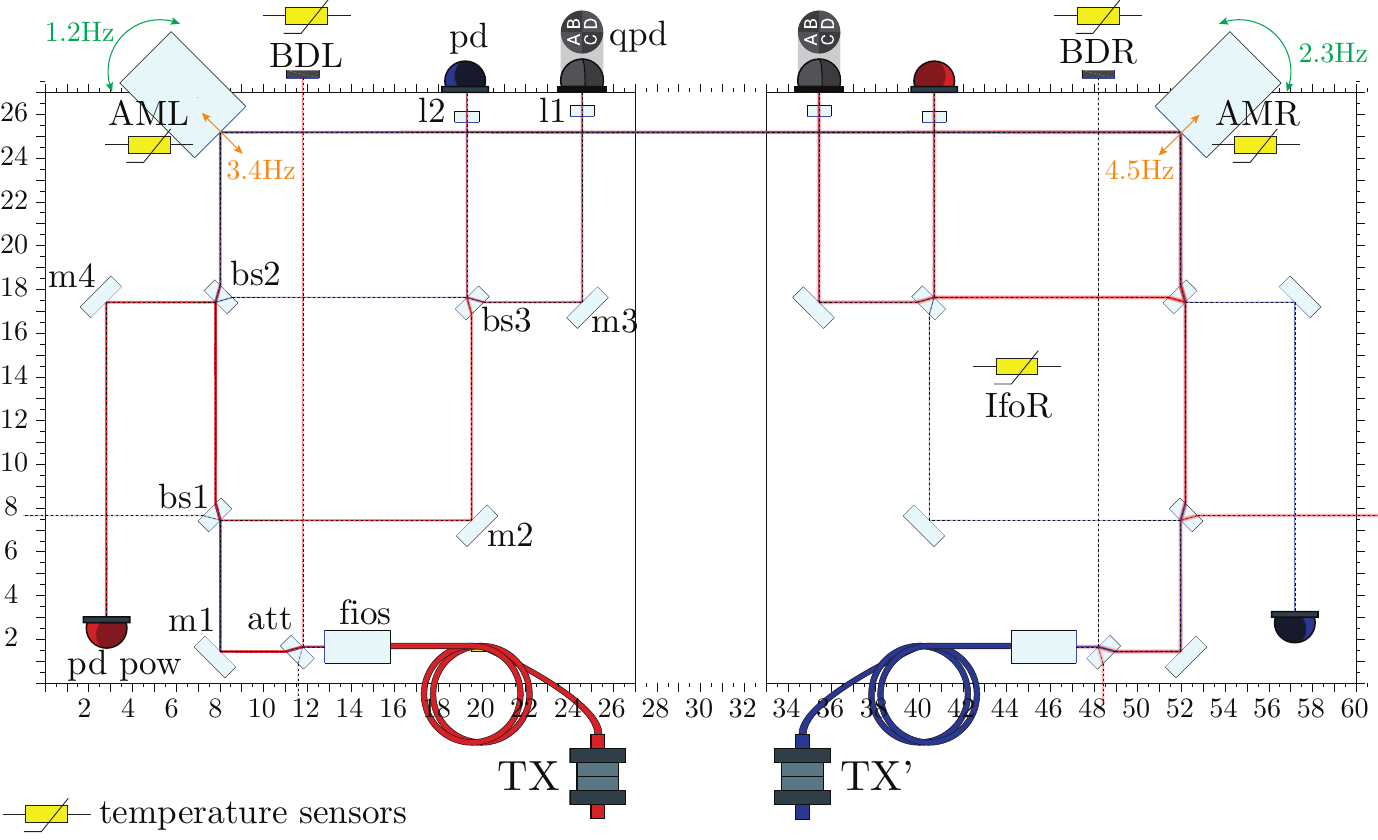}} \par \vspace*{0.15cm}
\subfigure[Photograph of the free beam link in the vacuum chamber inside the thermal shielding.]{\includegraphics[width=0.75\textwidth]{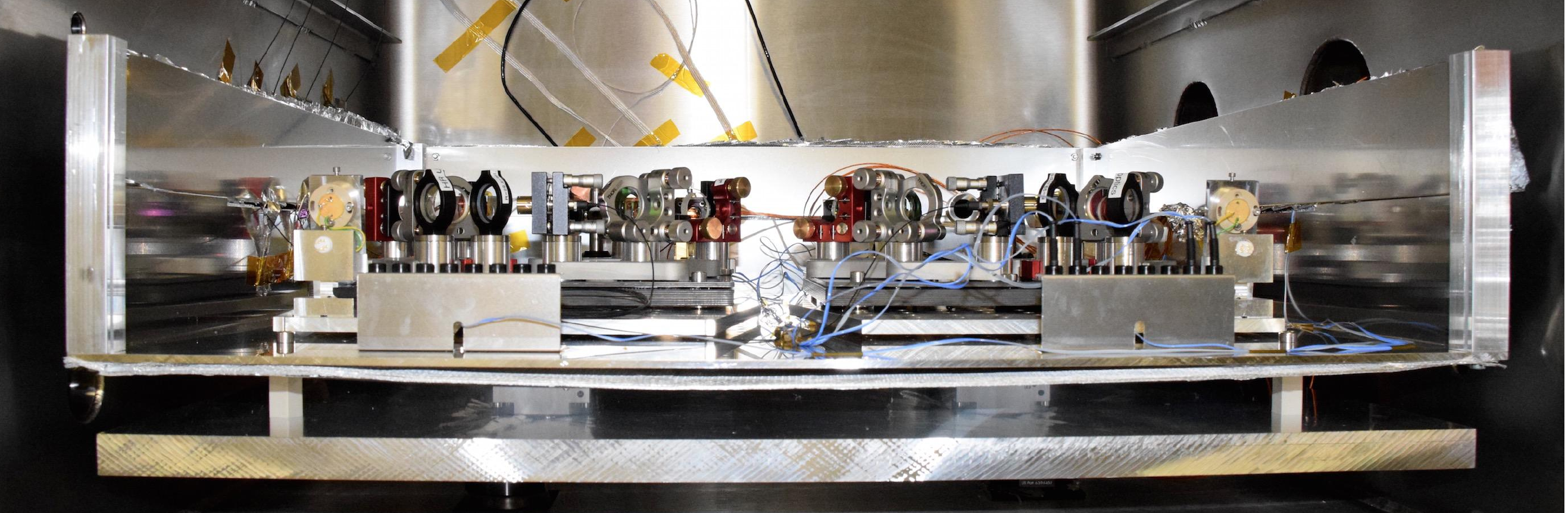}}

\caption{Layout (a) and photograph (b) of the experimental set-up of a free beam connection between two co-rotating benches. Off-the-shelf components and adjustable mounts are used on a carbon breadboard, placed on two aluminum plates sitting on two rotary tables. }
\label{fig:FreeBeam}
\end{figure*}

To test the thermal stability of our set-up during bench rotation and the implementation, operation and stability of the TBI free beam steering scheme we conducted an experiment with two optical benches realizing only this aspect of the TBI. The interferometer was also designed via IfoCAD and set up with off-the-shelf components and commercial fiber injectors placed on carbon breadboards, as shown in figure \ref{fig:FreeBeam}. The dimensions of the set-up are the same as the ones planned for the TBI, using the same steering mirrors and similar beam parameters with 1\,mm diameter and a waist position at the local steering mirror position. In contrast to the TBI, we did not implement polarization encoding, but used a stronger beam attenuation, 99\% instead of 95\% reflectivity, at the input FIOS. 
A description of the experimental set-up and the results are given in the following.


\subsection{Alignment of the free beam backlink}
The goal in aligning the free beam path is to collinearly overlap the inbound and outbound beams on each bench, with the steering control keeping this overlap constant. Non-uniform misalignments would otherwise lead to undesired tilt-to-pathlength noise during rotation. 
In order to achieve this alignment separately on each bench, we temporarily block the free beam link with an adjustable piezo mirror that reflects the light back onto the local bench. The mirror is aligned in such a way that it couples the reflected beam into the TX FIOS, making it collinear to the outbound beam. By applying a sinusoidal longitudinal motion to the mirror, the contrast in the interferometer can be observed and optimized by adjusting, for example, the recombining beam splitter. Finally, the temporary piezo mirror is removed and the two steering mirrors are aligned such that the contrast, that has previously been achieved purely with local beams, is recovered using the actual inter-bench beam combinations. This ensures the collinearity of the two opposing backlink beams. For the free beam experiment we achieve a contrast of more than 50\%.

\begin{figure}
  \subfigure[Time series of the motion of the rotary tables RTL (left) and RTR (right).]{  \includegraphics[]{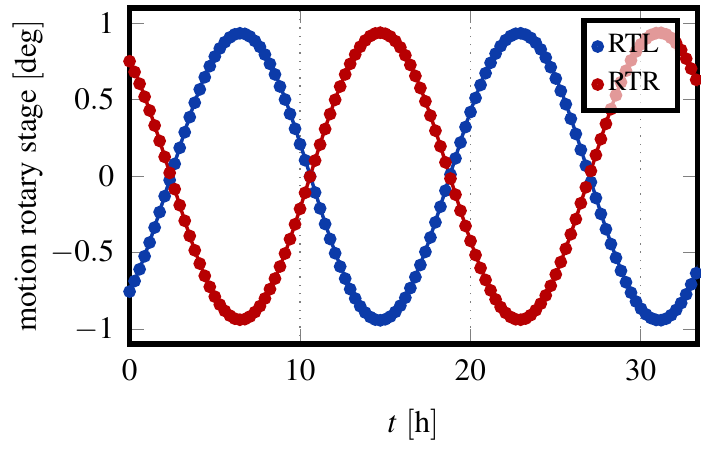}}  \hfill
    \subfigure[Time series of the vertical actuator mirror signals.]{  \includegraphics[]{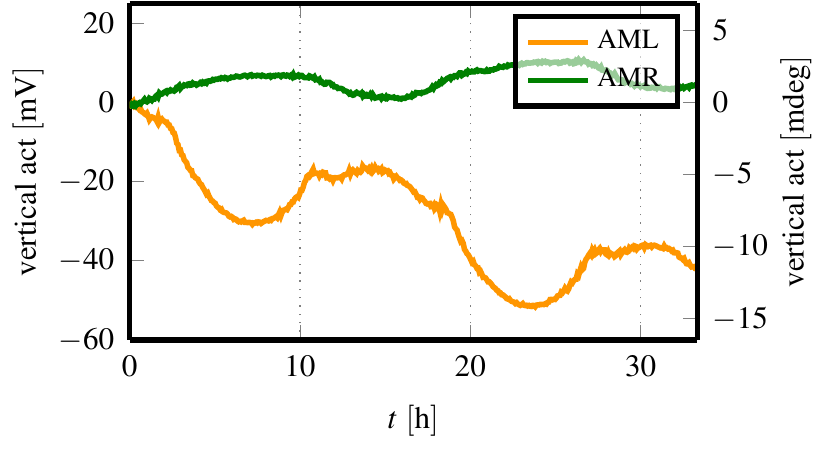}} 

  \subfigure[Time series of the horizontal DWS signals.]{  \includegraphics[]{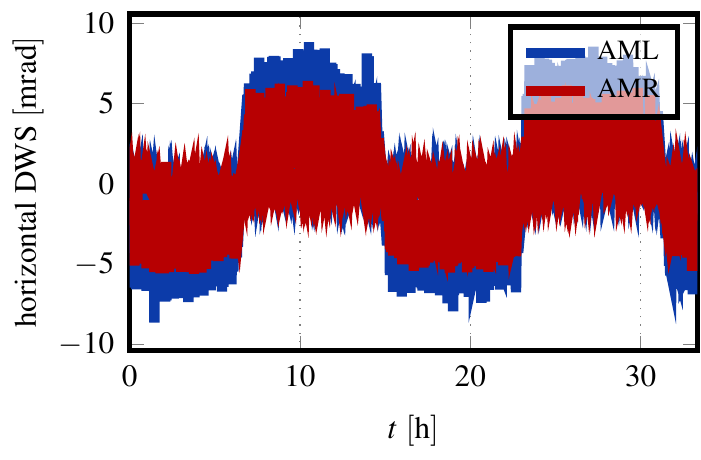}} \hfill
  \subfigure[Time series of the horizontal actuator mirror signals.]{  \includegraphics[]{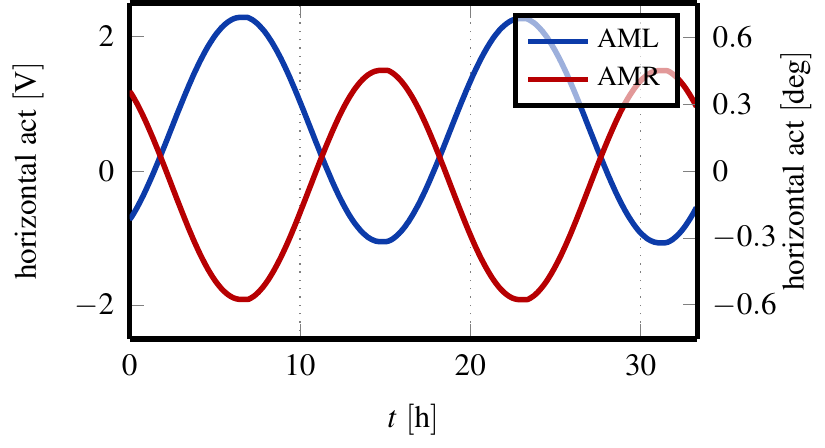}} 
	\caption{Time series of the rotary motion of the free beam experiment operating in vacuum. The two rotary tables actuate the interferometer benches with a period of $T_\mathrm{L} = 16.38\,\mathrm h$ and an amplitude of $0.9386\,\degree$. By monitoring the differential wavefront sensing (DWS) signals on both benches, 
	the error signals for the horizontal and vertical control loops of the actuator mirrors are derived by calculating the mirror misalignments from zero, $x_\mathrm l$, $x_\mathrm r$, $y_\mathrm l$ and $y_\mathrm r$.}
	\label{fig:DWStime}
\end{figure}

\subsection{DWS steering mirror control loop}
To keep the interferometers operational during rotation feedback control loops for each steering mirror axis are implemented. 
The control loops are coupled, so that a motion of one of the steering mirrors affects the phase readout on both sides. This behavior is described by the following coupling matrix: 

\begin{equation}
				\left(\begin{array}{c} \theta_\mathrm{{h l}} \\ \theta_\mathrm{{h r}}  \end{array}\right)
= \left(\begin{array}{cc} K1_\mathrm{hx} & K2_\mathrm{hx}\\ K3_\mathrm{hx} & K4_\mathrm{hx}   \end{array}\right)
\cdot 
				\left(\begin{array}{c} x_\mathrm{{l}} \\ x_\mathrm{{r}}  \end{array}\right)
\label{eq:matrix}
\end{equation}
with the horizontal DWS signals of the QPD on the left bench, $\theta_\mathrm{{hl}}$, its counterpart on the right bench, $\theta_\mathrm{{hr}}$, the coupling coefficients $K1-K4$ and the mirror motion in horizontal direction of the left steering mirror, $x_l$, and of the right steering mirror, $x_r$. An equivalent matrix is valid for the vertical coupling in relation to the vertical mirror motion $y_l$ and $y_r$. Due to misalignments of the QPDs and calibration errors of the steering mirror axes, a cross-coupling between horizontal and vertical motion and sensing is possible. In order to implement a control scheme for the free beam backlink equation (\ref{eq:matrix}) needs to be solved for the vector $(x_l, x_r)$. Thus the coupling matrix $K$ is required to be invertible and, beyond that, well-conditioned. 

A diagonal matrix represents a completely decoupled control system but is not achievable for this set-up. The coupling coefficients in vertical and horizontal direction are simultaneously measured by applying a sinusoidal motion to the steering mirrors with a different frequency in each axis, here 1.2\,Hz, 2.3\,Hz, 3.4\,Hz and 4.5\,Hz, as shown in figure \ref{fig:FreeBeam} (a). By demodulating the QPD response with the corresponding frequency, the influence of each mirror and each axis on the DWS signals is recovered. 
The coupling coefficients for the horizontal and vertical DWS control loops, calculated via equation (\ref{eq:matrix}), are:
\begin{equation}
	K_\mathrm{hx} =
	\left(\begin{array}{cc} 9323 & 4689\\ 5193 & 8794  \end{array}\right)
\,\frac{\mathrm{rad}}{\mathrm{rad}} \mathrm{;}\,\,  
	K_\mathrm{vy} =
		\left(\begin{array}{cc} 6860 & 3561\\ 3844 & 6476 \end{array}\right)
\,\frac{\mathrm{rad}}{\mathrm{rad}}.
\end{equation}
And for the cross-coupling we measured residual coefficients of
 \begin{equation}
	K_\mathrm{hy} =
			\left(\begin{array}{cc} 445 & 168\\ 92 & 33 \end{array}\right)
 \,\frac{\mathrm{rad}}{\mathrm{rad}} \mathrm{;}\,\,  
	K_\mathrm{vx} =
				\left(\begin{array}{cc} 132 & 24\\ 104 & 107 \end{array}\right)
 \,\frac{\mathrm{rad}}{\mathrm{rad}}.
\end{equation}

We achieve this decoupling by a factor of about 1.7 in both mirror axes by using imaging lenses with a focal length of 25.4\,mm in front of the photodiodes. These results are consistent with the optical simulations, which yielded a decoupling factor of 1.75 with coupling coefficients of 7000\,rad/rad and 4000\,rad/rad for a beam diameter of 1\,mm. In the simulation we observe different coupling coefficients depending on the distance between lens and photodiode, which was not further optimized in this set-up. This explains the deviation of experimental and simulated results as well as the deviation of the coupling coefficients between photodiodes. Ideally, the coupling coefficients should not change during rotation. Our simulation indicates a deviation of only 0.73\% for the decoupling factor for bench rotations of $+1\degree$ and $-1\degree$, experimental measurements show a worst case deviation of 5\% for $+0.75\degree$ and $-0.75\degree$. By inverting the measured matrices and forming linear combinations of the DWS signals, we calculate error signals for each mirror axis. This is then fed into integral controllers generating the actuation signals. We reach a unity gain frequency of 81\,Hz for the horizontal direction and 105\,Hz in the vertical, both exceeding a $30\degree$ phase margin. 

\subsection{Bench rotation}
The two rotary tables are rotated by $\pm 0.9\degree$ with a period of approximately 16 hours. The motion of both tables is done simultaneously with a relative delay of less than 0.02\,s limited by the controller accuracy. A time series measurement is depicted in figure \ref{fig:DWStime}(a). 
We achieve interferometer readout and steering mirror control for the full rotation period. The horizontal in-loop DWS signals and actuator signals are shown in figures \ref{fig:DWStime}(c) and (d). The vertical actuator signals are shown in figure \ref{fig:DWStime}(b). 

The jumps caused by the stepper motors in the rotary tables are clearly visible in the time series of the horizontal DWS signal in figure \ref{fig:DWStime}(c). The minimum incremental stepper motion of 0.0015\textdegree = 26\,\textmu rad corresponds to a path-length deviation of 4.3\,\textmu m per step on each bench in the free beam backlink path at the steering mirror position (163.45\,mm measured from the rotary axis). The antisymmetric bench rotation produces a total accumulated path-length for two simultaneous steps of 8.6\,\textmu m, which corresponds to almost 9 cycles considering a wavelength of 1064\,nm. 
 By integrating eddy current dampers in to the mechanical set-up, the resonance frequency is being damped. 

\begin{figure*}
\centering
  \includegraphics[]{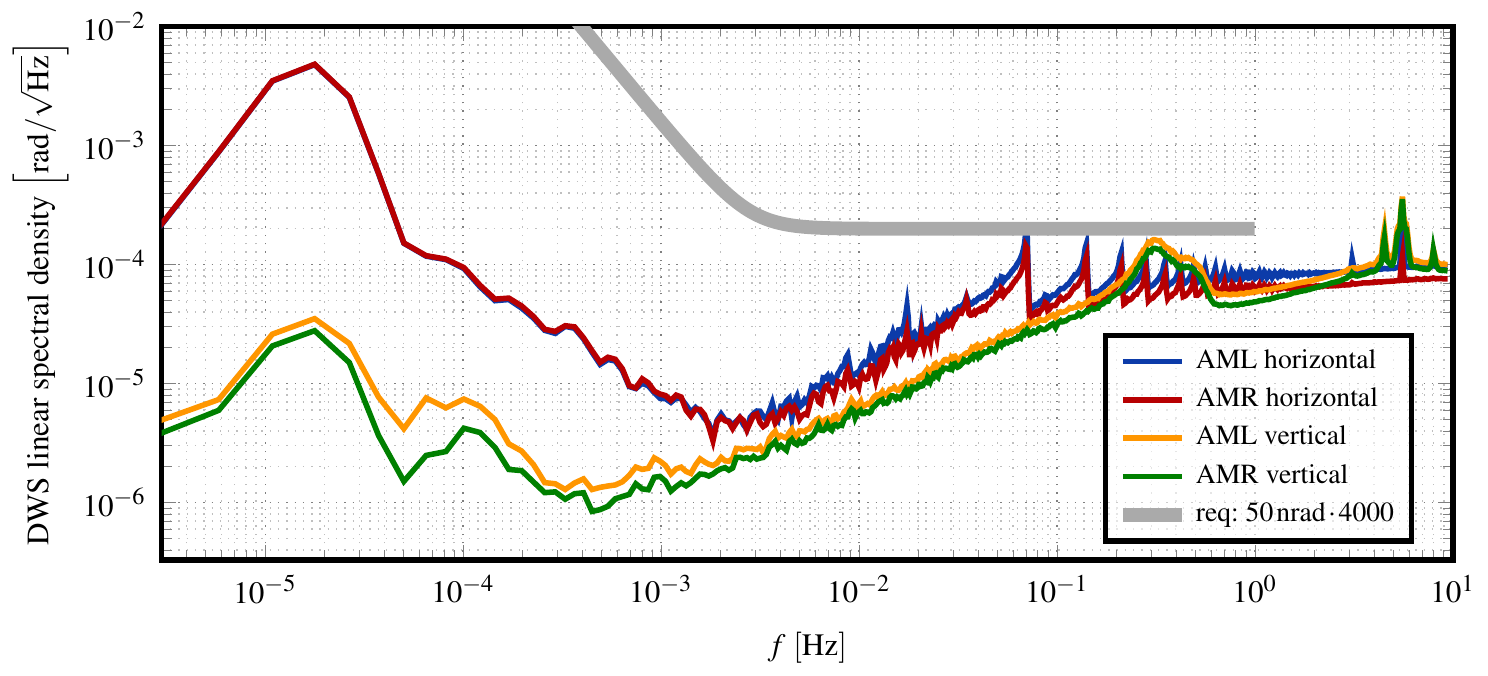}
  
\caption{Linear spectral densities of the differential wavefront sensing signals corresponding to the left and right actuator mirror (AML and AMR) and both directions, horizontal (figure \ref{fig:DWStime} (c)) and vertical. The plotted requirement is derived in section 2 and corresponds to 0.2\,mrad DWS accuracy, with the assumption that the coupling coefficient from phase to DWS is at least 4000\,rad/rad.}
\label{fig:DWSresult}
\end{figure*}

Figure \ref{fig:DWSresult} shows the linear spectral densities of the measured DWS signals for the horizontal and vertical directions. Here the jumps are visible as peaks between 0.07\,Hz and 1\,Hz. As explained in section 2 the requirement for the residual angular jitter is $0.2\,\milli\radian/\sqrt{\hertz}$, which is only just achieved with the current control.

\subsection{Long term measurements}
The second output port of the recombination beam splitters enables us to perform two measurements of the longitudinal phase on each bench. We use a single-element photodiode (SED) to implement an out-of-loop measurement that can also be used for diagnostic purposes. In total, we have five phase signals per bench, the SED phase signal, $\varphi^p_\mathrm{SED}$, and the phase signals of the four segments of the QPD, $\varphi^p_\mathrm{A}$, $\varphi^p_\mathrm{B}$, $\varphi^p_\mathrm{C}$ and $\varphi^p_\mathrm{D}$, with $p = \{\mathrm{left}, \mathrm{right} \}$. The QPD phases can be combined into a single phase via
\begin{equation}
	\varphi^p_\mathrm{QPD} = \frac{\varphi^p_\mathrm{A}+\varphi^p_\mathrm{B}+\varphi^p_\mathrm{C}+\varphi^p_\mathrm{D}}{4}.
\end{equation}
The investigation of noise sources in the current experimental set-up makes use of different combinations. The so-called $\pi$-measurement compares the two output ports of a single recombination beam splitter. The difference of the corresponding phase signals, 
\begin{equation}
	\varphi^p_\pi = \varphi^p_\mathrm{SED} - \varphi^p_\mathrm{QPD} \approx \pi, 
\end{equation}
theoretically reduces to a constant phase shift of $\pi$. For the investigation of the backlink connection, the signals of both benches are combined by calculating the phase sum of the two benches, 
\begin{equation}
	\varepsilon_\mathrm{SED} = \varphi^\mathrm{right}_\mathrm{SED} + \varphi^\mathrm{left}_\mathrm{SED} =\varphi_\mathrm{NR} + 2\varphi_\mathrm{CPN}.
\end{equation}
This combination contains the non-reciprocity, $\varphi_\mathrm{NR}$, and a common phase noise term, $\varphi_\mathrm{CPN}$.
A direct measurement of the non-reciprocity would require a reference signal to subtract the common phase noise term from $\varepsilon_\mathrm{SED}$. However, this has not been implemented in our experiment. The last combination is the difference of the left and right benches' phase signals
\begin{equation}
	l_\mathrm{SED} = \varphi^\mathrm{right}_\mathrm{SED} - \varphi^\mathrm{left}_\mathrm{SED},
\end{equation}
which directly corresponds to twice the length of the backlink path and includes laser frequency noise. 

\begin{figure*}
\centering
  \includegraphics[]{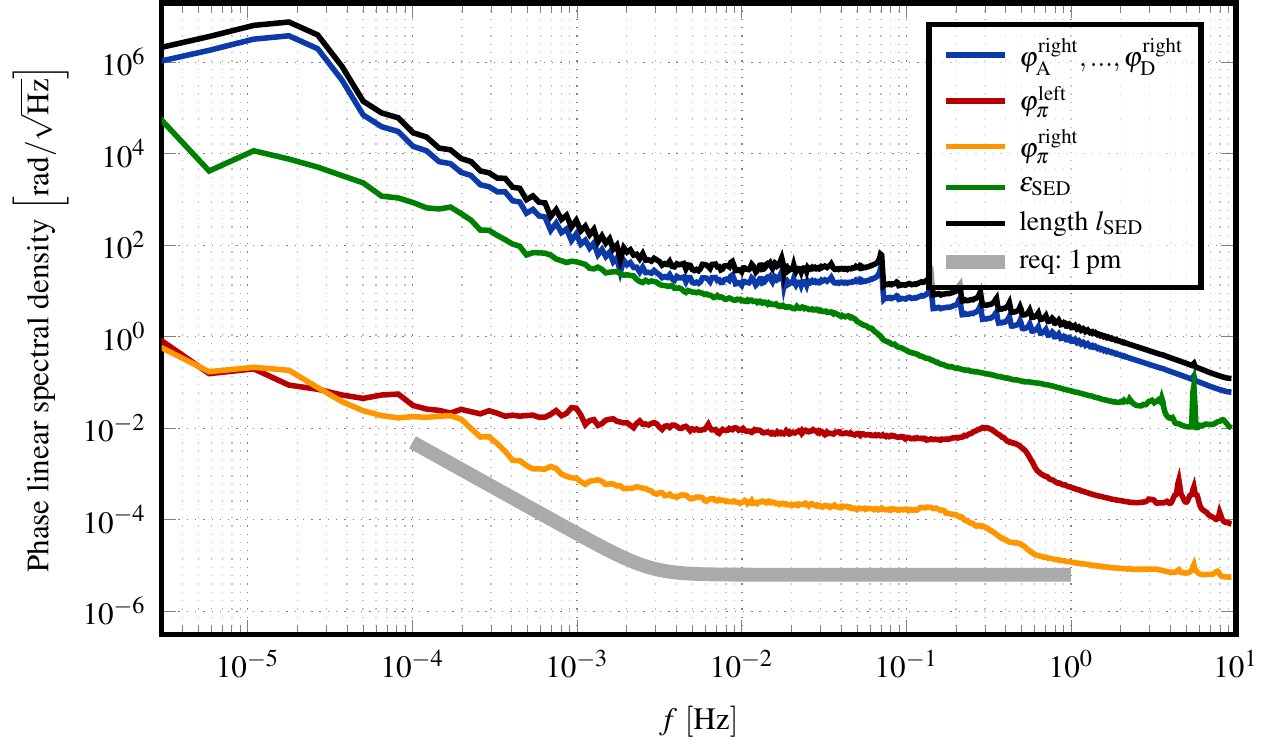}
\caption{Linear spectral densities of phases measured via various photo detector segments, $\varphi_\mathrm A^\mathrm{right}$, ... $\varphi_\mathrm D^\mathrm{right}$, and combinations thereof. The $\pi$-measurements, $\varphi^\mathrm{left}_\mathrm{\pi}$ and $\varphi^\mathrm{right}_\mathrm{\pi}$, compare the phases of the two output ports of the recombination beam splitter on the left and right bench. The sum, $\varepsilon_\mathrm{SED}$, shows an upper limit for the non-reciprocity. In contrast, the difference of the left and right phase measurements provides an absolute length measurement of the backlink path, $l_\mathrm{SED}$, including laser frequency noise.}
\label{fig:phaseresult}
\end{figure*}

Figure \ref{fig:phaseresult} shows the linear spectral densities of the measured phases along with their relevant combinations. 
All photo detectors and segments thereof show a similar noise behavior that is given by the blue curve. For this reason, the aforementioned signal combinations are a good diagnostic tool for understanding the various noise sources. 
For the $\pi$-measurements (red and orange curves), we observe a typical stray light shoulder in both spectra with different noise levels. Additional ghost beams on the left photodiode lead to a shoulder noise level of $10^{-2}\,\radian/\sqrt{\hertz}$ between 1\,mHz and 100\,mHz, while the photodiode on the right shows almost 2 orders of magnitude less noise in its phase measurement. However, the influence of the ghost beams or scattered light on the phase depends on the position of the rotation tables and varies throughout the rotation. A detailed analysis of the time series (not depicted) shows that the absolute phase of the ghost beam increases to $15\,\milli\radian$ during each semi-period and becomes minimal at the reversal points. Without motion the phase fluctuations of the ghost beam are in the range of $3\,\milli\radian$. The right interferometer contains a ghost beam with a phase magnitude of $0.75\,\milli\radian$. 

The path-length change due to the rotation is proportional to the motion itself, which was to be expected. This is shown in the spectrum of the path length (black curve), which shows the same oscillation pattern as the spectrum of the horizontal DWS from figure \ref{fig:DWSresult}. 

The phase combination, $\varepsilon_\mathrm{SED}$, exhibits a 1/$f$ decrease in the spectral density with  $65\,\milli\radian/\sqrt{\hertz}$ at 1\,Hz, which is equivalent to $16\,\mathrm{nm}/\sqrt{\hertz}$ at 1\,Hz. 
A missing local phase reference means that this measurement does, as expected, not achieve the requirement and that this is only possible with a TBI-like experiment. The excess noise is driven by  phase noise from the laser preparation in air and two 12\,m long fibers. In the TBI the direct fiber backlink will be used as local reference with $1\,\mathrm{pm/\sqrt{Hz}}$ non-reciprocity, enabling us to measure the phase noise behavior of the other phase references down to the same level. Furthermore, as described in section 2, the TBI's performance will improve through ghost beam suppression, usage of a quasi-monolithic set-up, polarization cleaning, power stabilization and potentially fiber length stabilization.

\subsection{Thermal measurement results}

\begin{figure}
\centering
\subfigure[Time series of the temperatures inside the vacuum chamber with moving rotary tables and at rest in vacuum.]{  \includegraphics[]{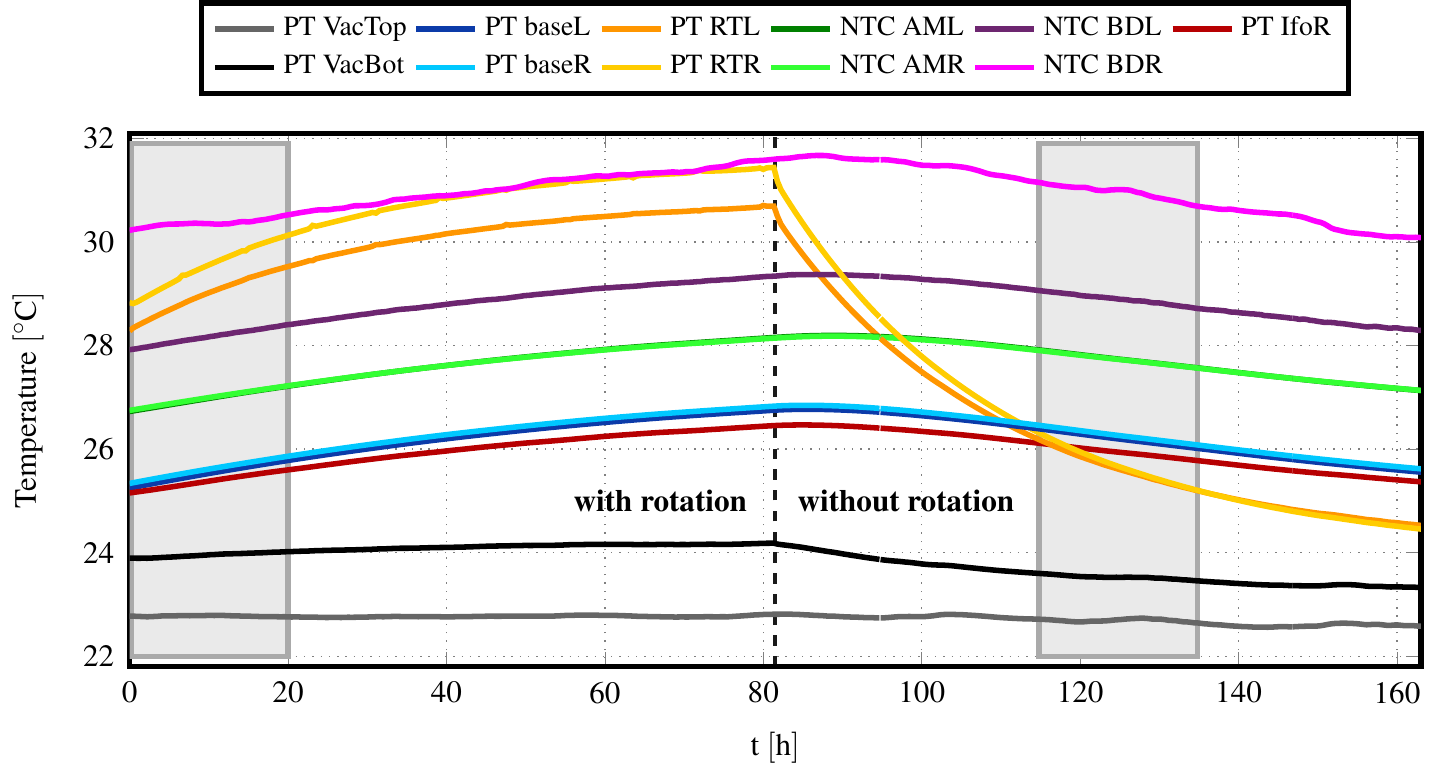}} 
  
  \subfigure[Temperature noise with moving rotary tables.]{  \includegraphics[]{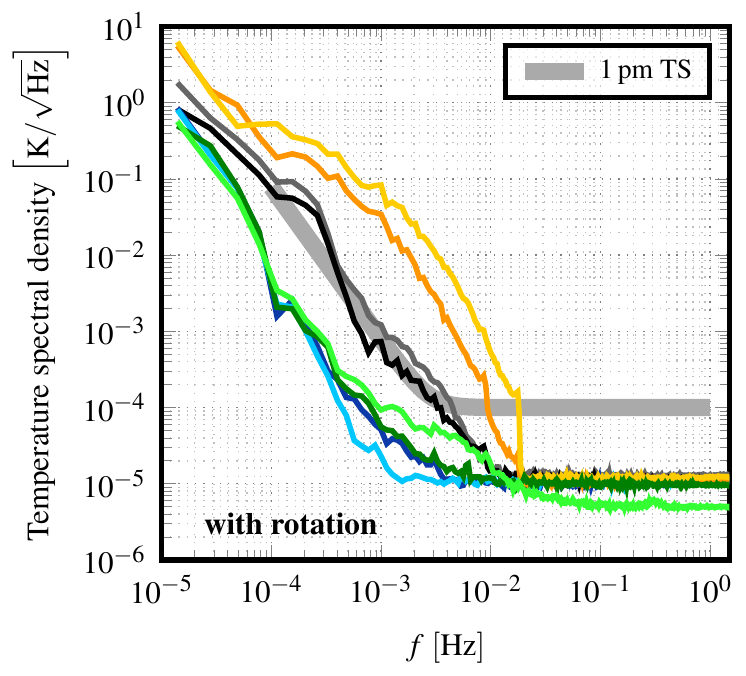}} \quad
  \subfigure[Temperature noise with rotary tables at rest.]{  \includegraphics[]{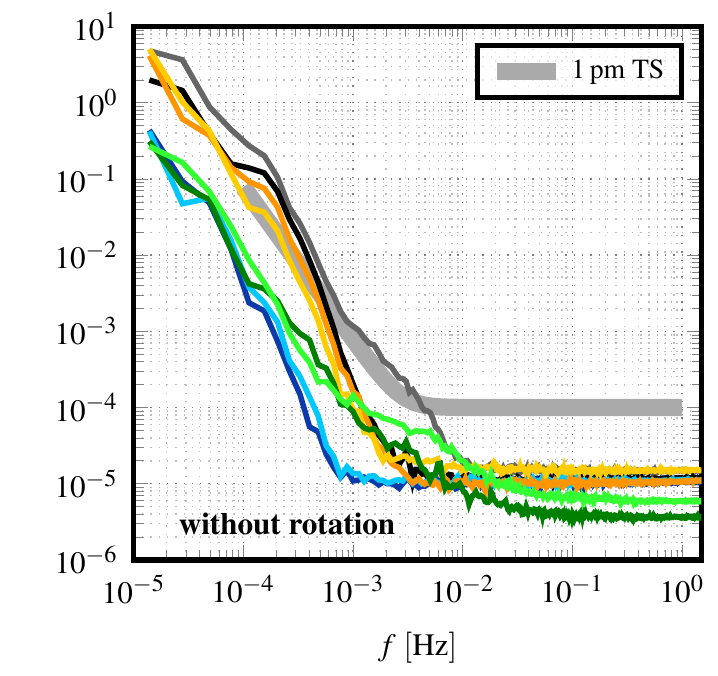}} 
\caption{Different temperature sensors, namely platinum resistance temperature sensors PT10000 and negative temperature coefficient thermistors (NTCs), monitor various locations in the vacuum chamber, as shown in figures \ref{fig:FreeBeam} (a) and \ref{fig:PreStab} (b). The lower two plots show the linear spectral densities of the temperatures for the time intervals highlighted in the time series plot above for selected temperature sensors. The requirement for the thermal stability (TS) is derived from a 1\,pm displacement stability in a quasi-monolithic interferometer with an absolute arm-length mismatch of 1\,m, using a baseplate with a coefficient of thermal expansion of $1\cdot 10^{-8}/\kelvin$.}
\label{fig:tempresult}
\end{figure}

To verify the thermal stabilities we read out the temperature sensors at critical points throughout the experiment. Figures \ref{fig:FreeBeam}(a) and \ref{fig:PreStab}(c) show the positions of a selected few that are relevant for the analysis, while our measurement results are compiled in figure \ref{fig:tempresult}. The time series plot in the latter shows both, the temperature behavior during rotation of the interferometers while actuating on the steering mirrors for 80 hours, and the pronounced temperature drop for the rotary tables at rest during the following 80 hours. In both cases the actuator mirrors are supplied with control voltages, but, for the second case, these turn out to be almost constant. The linear spectral density graphs below the time series plot display the temperature noise for both cases. The highlighted areas in the time series mark the sections used for the calculation of the spectra.

The increase of the absolute temperatures during rotation, which can be observed in the first half of the time series, was to be expected. This effect is very significant for the rotary tables themselves, which reach a maximum temperature of $31.4\,\celsius$ in this measurement, corresponding to an increase of $2.6\,\celsius$ over 80 hours. The highest temperature, namely $31.6\,\celsius$, is measured at one of the beam dumps, while the most stable temperature could be observed at the ceiling plate inside the vacuum chamber. Once the motion of the tables is completed the temperatures start to decrease. Again, this
is most significant for the sensors on the rotary tables. They detect a decrease of $7\,\celsius$ down to $24.4\,\celsius$.
With the temperature, also the noise levels of the temperature of the rotary tables increases for frequencies 
up to $2\cdot 10^{-2}\,\hertz$, while the noise level of the actuator mirrors stay mostly stable. 
The most relevant temperature behavior with regard to the TBI is the thermal stability of the two rotating baseplates (baseL and baseR), carrying the interferometric benches. For both cases, with and without rotation, the two relevant sensors (blue curves) attest to a sufficiently low temperature noise of $10^{-4}\,\kelvin/\sqrt{\hertz}$ at 1\,mHz. 
The white noise level for all sensors is at about $10^{-5}\,\kelvin/\sqrt{\hertz}$ and is reached somewhere between 2.2\,mHz and 18\,mHz.

\section{Conclusion and Outlook}
LISA requires an optical phase reference between the two optical benches in each satellite that has a non-reciprocity below $1\,\mathrm{pm}/\sqrt{\hertz}$. For these phase references the ghost beams generated by fiber back scatter are critical, as is the case for the performance of all local interferometers. We have presented the detailed design of an experiment that will compare two alternative implementations of this phase reference to a previously tested direct fiber link. This will give us detailed insight into the noise limitations for each. The experimental infrastructure required to operate this interferometer has been set up and tested together with the steering mirror control for a free beam phase reference scheme. As a next step we will construct the Three-Backlink Interferometer, integrate it into our experiment and begin the noise hunting process. 

The alternative schemes we will investigate have the potential to be less susceptible to fiber back scatter, promising lower noise levels and/or hardware requirements better-suited for the overall mission. Balanced detection, which puts additional constraints on the redundancy and readout channel numbers in LISA, is unavoidable if a direct fiber link is chosen. In our current understanding, the alternative schemes will not require this correction or require it to a much lower degree. However, a direct, experimental comparison of all three schemes within the Three-Backlink Interferometer will be necessary to validate our noise assumptions. These results, together with other LISA breadboarding experiments, shall, in the end, inform the study to find the best overall solution for LISA.

\section*{Acknowledgement}
The authors would like to thank the DFG Sonderforschungsbereich (SFB) 1128 Relativistic Geodesy and Gravimetry with Quantum Sensors (geo-Q) for financial support. We also acknowledge support by the Deutsches Zentrum f\"ur Luft- und Raumfahrt (DLR) with funding from the Bundesministerium f\"ur Wirtschaft und Technologie (Project Ref. No. 50 OQ 0601) and the European Space Agency (ESA) within the project Phase reference distribution system (8586/16/NL/BW). 

We also would like to thank Guido M\"uller from the University of Florida, Jeffrey Livas from the NASA's Goddard Space Flight Center, Ewan Fitzsimons from the Astronomy Technology Centre in the UK and Henry Ward, David Robertson, Christian Killow and Michael Perreur-Lloyd from the University of Glasgow in the UK for helpful discussions.

\section*{References}
\providecommand{\newblock}{}

\bibliographystyle{iopart-num}

\end{document}